\documentclass[aip,rsi,preprint,superscriptaddress]{revtex4-1}
\usepackage[hypertexnames=false]{hyperref}
\hypersetup{colorlinks=true,citecolor=black,linkcolor=black,urlcolor=black}

\AtBeginDocument{
\renewcommand{\ref}[1]{\autoref{#1}}

}
\usepackage{amsfonts}
\usepackage{graphicx}
\usepackage{lineno}
\usepackage[usenames]{color}

\begin{document}

\title{A versatile rotary-stage high frequency probe station for studying
magnetic films and devices \smallskip{}
}

\author{Shikun He}

\affiliation{Data Storage Institute, A{*}STAR (Agency for Science, Technology
and Research), 2 Fusionopolis Way, \#08-01, Innovis 138634, Singapore}

\affiliation{Division of Physics and Applied Physics, School of Physical and Mathematical
Sciences, Nanyang Technological University, Singapore 637371}

\author{Zhaoliang Meng}

\affiliation{Data Storage Institute, A{*}STAR (Agency for Science, Technology
and Research), 2 Fusionopolis Way, \#08-01, Innovis 138634, Singapore}

\author{Lisen Huang }

\affiliation{Data Storage Institute, A{*}STAR (Agency for Science, Technology
and Research), 2 Fusionopolis Way, \#08-01, Innovis 138634, Singapore}

\author{Lee Koon Yap}

\affiliation{Data Storage Institute, A{*}STAR (Agency for Science, Technology
and Research), 2 Fusionopolis Way, \#08-01, Innovis 138634, Singapore}

\author{Tiejun Zhou}

\affiliation{Data Storage Institute, A{*}STAR (Agency for Science, Technology
and Research), 2 Fusionopolis Way, \#08-01, Innovis 138634, Singapore}

\author{Christos Panagopoulos}

\affiliation{Division of Physics and Applied Physics, School of Physical and Mathematical
Sciences, Nanyang Technological University, Singapore 637371}
\begin{abstract}
We present a rotary-stage microwave probe station suitable for magnetic
films and spintronic devices. Two stages, one for field rotation from
parallel to perpendicular to the sample plane (Out-of-Plane) and the
other intended for field rotation within the sample plane (In-Plane)
have been designed. The sample, probes and micro positioners are rotated
simultaneously with the stages, which allows the field orientation
to cover $\theta$ from 0 to $90^{\circ}$ and $\varphi$ from 0 to
$360^{\circ}$. $\theta$ and $\varphi$ being the angle between the
direction of current flow and field in a Out-of-Plane and an In-Plane
rotation, respectively. The operation frequency is up to 40 GHz and
the magnetic field up to 1 T. The sample holder, vision system and
probe assembly are compactly designed for the probes to land on a
wafer with diameter up to 3$\,$cm. Using homemade multi-pin probes
and commercially available high frequency probes, several applications
including 4-probe DC measurements, the determination of domain wall
velocity and spin transfer torque ferromagnetic resonance are demonstrated. 
\end{abstract}
\maketitle

\section{introduction}

The discovery of spin transfer torque has attracted considerable attention
in the past decade.\citep{Slonczewski_STT_1996,Mangin_STT_TMR,Rippard_PRL_STO,NatM_STO_2007,Ikeda_2010_Namat}
Recently, manipulation of magnetic bits through spin Hall effect\citep{liu_SHE_CFB}
and spin orbit torque\citep{YuGQ_SOT_2014} have also been demonstrated.
The characteristic time for current induced magnetic switching is
ns and the typical frequency of spin torque driving response is in
the GHz range. These findings have made magnetic nanodevices realistic
candidates for memory and microwave device applications. In response,
there has been increasing demand for microwave frequency probe stations
capable of characterizing magnetic films and devices.

Various probe stations with variable temperature and magnetic field
are commercially available. (Lake Shore Cryotronics, Inc. and Janis
Research Company for example.) However, for room temperature applications,
it is imperative to have the option of changing field direction with
respect to the sample. This is especially important because magnetic
materials are generally anisotropic and the effect of spin torque
is angle dependent. Unfortunately, only a few solutions for the field
rotation probe station are available. Technically, one needs to either
rotate the magnet system or the sample. The former is favorable when
signal stability is a major concern. Homemade vector magnets and projected
field electromagnets (GMW associates) can provide a rotating magnetic
field, however, the field uniformity is poor and the field strength
is usually limited to a few kOe.\citep{GMW_project} As detailed in
a recent design,\citep{vectorMagnet2012} extra care must be taken
to calibrate the magnets. Besides, close loop PID control is much
more complicated for applications require precise field values. A
magnet with rotating base, on the other hand, only enables the field
rotation along a single axis. In the latter case, the commercial products
have adopted sample stage rotation design to allow 360$^{\circ}$
of sample rotation within the film plane.\citep{Lakeshore_PS,Janis}
The orientation of the probes and the corresponding positioners are
fixed. For devices working at low frequency ($<$1$\,$GHz), the DC probes
can always be landed on the pads without any issue. However, in a
high frequency ($\sim$10$\,$GHz) measurement, impedance match is 
of primary importance. Consequently, special probes with integrated
tips such as ground-signal-ground (GSG) probes and coplanar waveguide
type device pads should be used.\citep{nature_TI_2014} The misalignment
between the probe and the pads of the device is usually limited to
30$^{\circ}$ for reliable measurement.\citep{wangyi_APL} Several
devices with identical pattern but different orientations must be
measured to obtain device parameters as a function of field direction
in the range of 0 to 360$^{\circ}$. In reality, geometry as well
as magnetic property variations among sub-100$\,$nm devices are inevitable.
Furthermore, re-landing of the probes to the device is necessary for
each sample orientation, which greatly increases the testing time,
causes reliability issue and hinders fully automation.

Here, we present a compact and systematic solution for building a
RF probe station with relative field orientation capable of rotating
from parallel to perpendicular to the sample plane (out-of-plane (OP)
rotation) as well as 360$^{\circ}$ rotation within the sample plane
(in-plane (IP) rotation). The design is based on the idea of rotating
sample, probes and microwave cables simultaneously, making the angular
dependent measurement convenient and efficient. The test platform
is suitable for studying magnetic films and devices from DC to at
least 40 GHz. We also demonstrate examples of the capabilities of
the apparatus through magnetoresistance (MR) and angular magnetoresistance
(AMR ) measurements on a film and two devices, the determination of
domain wall velocity and a test of spin-transfer-torque ferromagnetic
resonance (STT-FMR).

\section{apparatus}

\subsection{Magnet system}

\begin{center}
\begin{figure*}
\centering{}\includegraphics[width=0.8\textwidth]{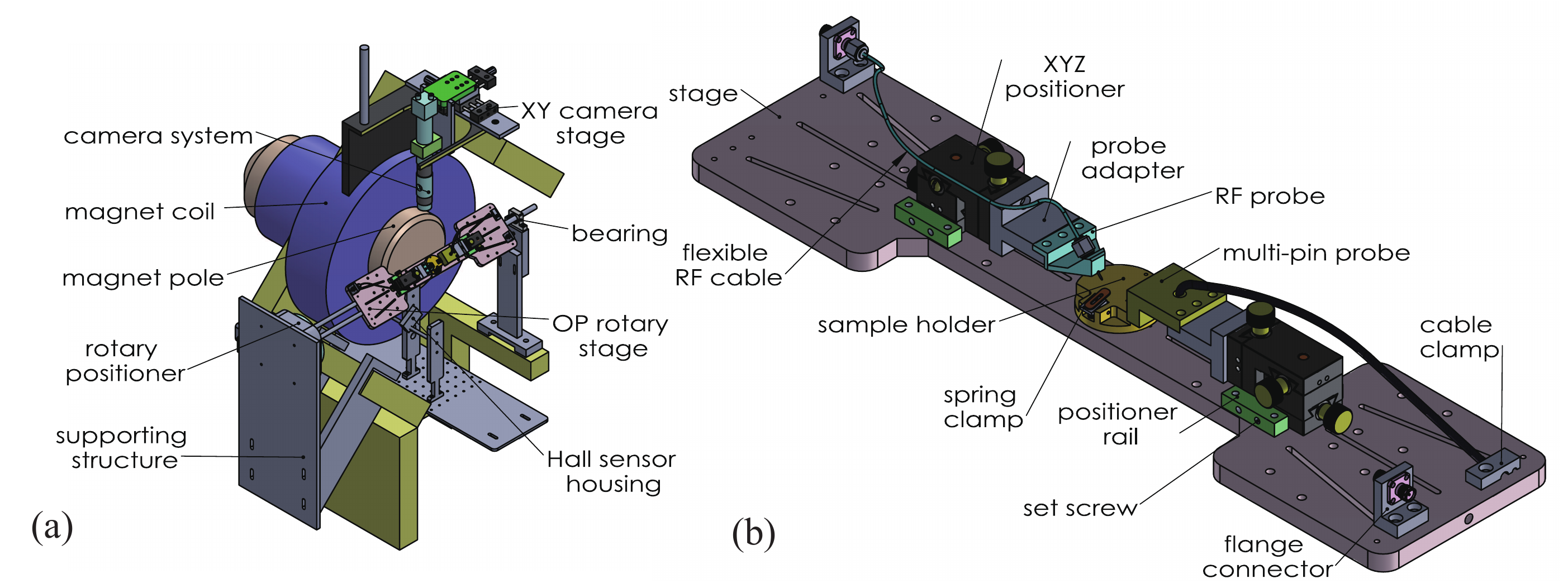}\protect\caption{(a) Illustration of the RF probe station with OP rotary stage mounted.
(b) Zoom in view of the OP rotary stage.\label{fig:OP_stage}}
\end{figure*}

\par\end{center}

The electromagnet used for this probe station was purchased from Lake
Shore Cryotronics, Inc.. A pair of poles with 15.2$\,$cm face diameter
were selected. The position of the poles is adjustable with a maximum
gap of 17.8 cm. For a pole gap of 10 cm, the field uniformity at the
center is better than $\pm$0.5\% over a sphere of 2.5$\,$cm diameter.
The samples are placed approximately at the center of the sample holder
to further minimize the field discrepancy during sample rotation.
The magnet power supply with a current range of ±135 A is controlled
by Model 475 Gaussmeter. Magnetic field as large as 1 T can be precisely
applied. Our probe station described below is configured for this
magnet, however, the method can be applied to other commercially available
magnets as well.

\subsection{Probe station with OP rotary stage}

A schematic of the setup with the OP rotary stage installed is shown
in \ref{fig:OP_stage} (a). The probe station consists of three main
parts: the above mentioned magnet, a vision system and a stage assembly
driven by a motorized rotary positioner. Auxiliary parts including
supporting structures for the stage and Hall sensor mounting kit are
also necessary. The vision system rests on a XY positioner in order
to zoom in a device anywhere in a circle of 5$\,$cm diameter. The
camera and the lens provide a maximum total magnification of 70. The
heavy duty rotary positioner (RM-5 Newmark systems, INC.) has a maximum
output torque of 10.2 N$\cdotp$m. One end of the stage was mounted
on the positioner and the other end with a guiding tube ran through
a bearing. With this configuration, the stage can be oriented with
an accuracy of $0.02{}^{\circ}$. 

\ref{fig:OP_stage} (b) is a closer view of the OP stage. The sample
holder with a diameter of 4 cm is equipped with a spring-loaded clip
for easy and reliable sample mounting as well as sample position adjustment.
The springs are made of BeCu to avoid magnetic impurities. Due to
limited space, we used low profile non-magnetic XYZ micro positioners
(Quater Research and Development) which have a travel distance of
0.75$\,$cm along each axis. The positioners are partially exposed
to the magnetic field. Instead of using vacuum and magnetic mounting
base, our positioners were mounted on the stage through rails and
locked by set screws. This method improves the stability and allows
an additional position adjustment of 2$\,$cm. Up to 4 positioners
can be mounted to the stage, which enables a standard four probe low
frequency measurement. As can be seen from \ref{fig:OP_stage} (b),
probe adapters were used to secure the preferred RF probes and the
homemade multi-pin probe. Nonmagnetic flexible RF cables with appropriate
length were used. To minimize the vibration of probes due to cable
bending, flange connector assemblies and cable clamp at the edge of
the stage is crucial. Note that we have minimized the usage of magnetic
components for the stage.

\begin{figure}
\centering{}\includegraphics[width=8cm]{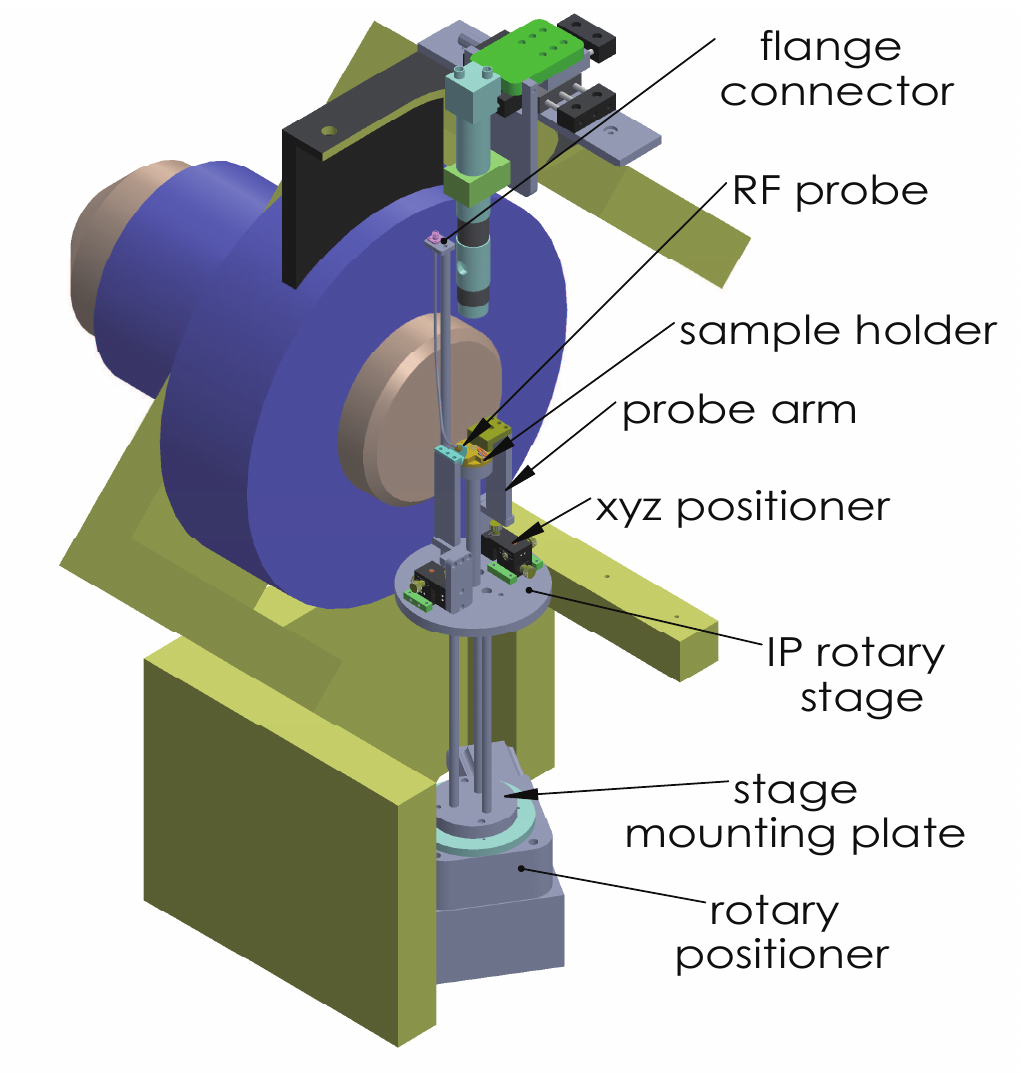}\protect\caption{Schematic of the probe station with IP rotary stage mounted. \label{fig:IPR}}
\end{figure}

The home position of the stage was set for the sample surface parallel
to the applied magnetic field. ($\theta_{H}$=0 as illustrated in
\ref{fig:illus_rotation} (a).) It is necessary to land the probes
on the pads (with the guidance of the vision system) at the home position
before performing an actual experiment.

\subsection{IP rotary stage}

The probe station with the IP rotary stage mounted is shown in \ref{fig:IPR}.
The sample holder placed at the field center, the XYZ micro positioners
mounted on the stage and the probes are identical to the ones used
in the previous OP stage. The rotary positioner was mounted on a heavy
base and the axis of rotation (Z) was aligned to the field center.
The circular shape rotary stage, at a height convenient for operating
the micro positioners, was mounted through three stainless steel 316
tubes. The probe arm assemblies made of Polyether ether ketone (PEEK)
are of essential to reduce the pole gap. Here, with two positioners
mounted, the minimum pole gap is 7.5$\,$cm in order to perform an
IP rotation. Whenever a RF probe is used, the other end of the RF
cable is connected to a RF flange connector fixed at the top of an
Al tube. Note that a right angle RF probe (with connector along Z
axis which can further reduce the pole gap) is more favorable than
the 45 degree probe illustrated in \ref{fig:IPR}.

\section{Experimental test and applications}

\begin{figure}
\centering{}\includegraphics[width=7.8cm]{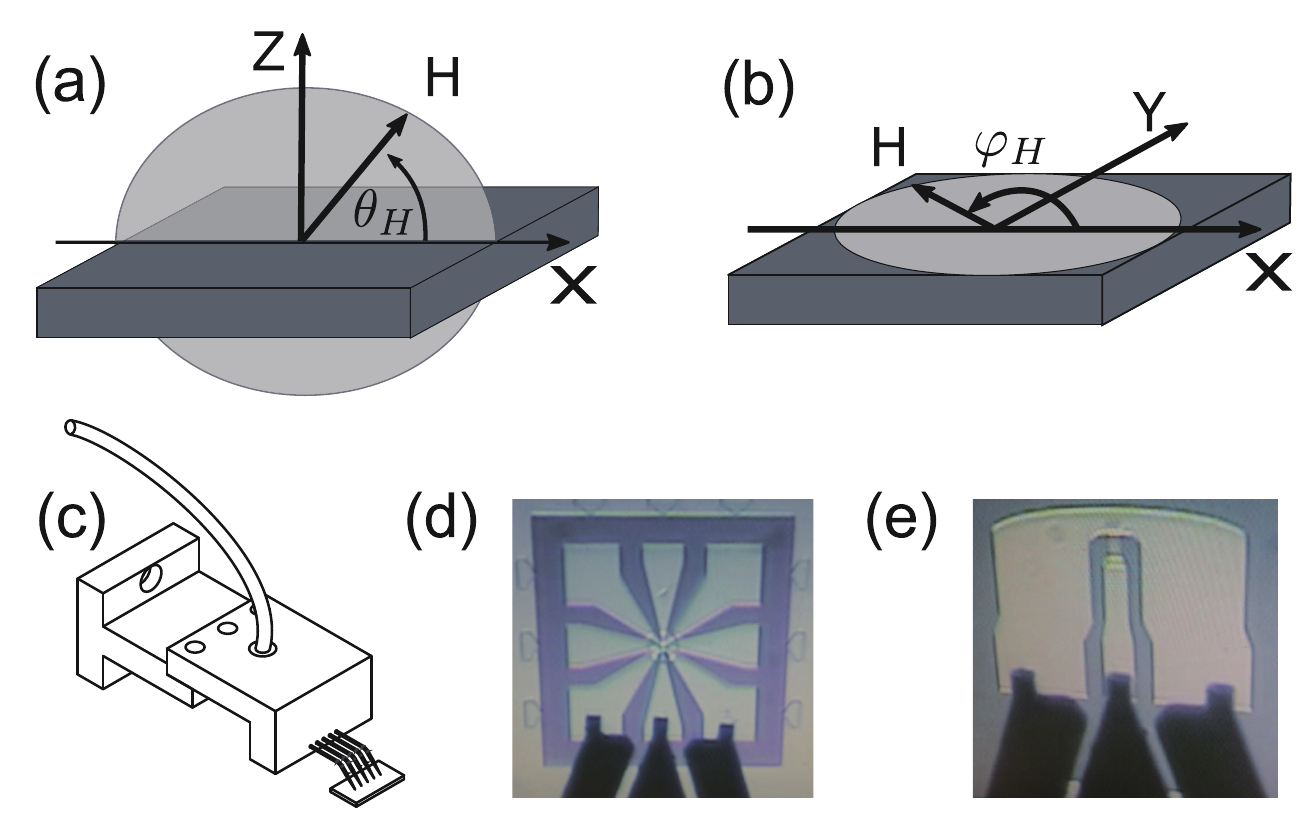}\protect\caption{Illustrations of the measurement configurations using the RF probe
station. The stage rotation is equivalent to a field rotation in the
sample coordinate system for OP stage (a) and IP stage (b). $\theta_{H}$
and $\varphi_{H}$ are defined as the angle between field and the
direction of current flow. (c) The film level measurement using homemade
multi-pin probe. (d) GSG probe landed on a magnetic tunnel junction
device. (e) GSG probe landed on a coplanar waveguide which is connected
to a stripe device. \label{fig:illus_rotation}}
\end{figure}

In this section, we present several sets of experimental data to demonstrate
the capability of the apparatus. The coordinate system of the probe
station is schematically shown in \ref{fig:illus_rotation} (a) and
(b). $\theta_{H}$ is the angle between magnetic field and the sample
film plane in a OP rotation whereas $\varphi_{H}$ corresponds to
the angle between the magnetic field and the direction of current
flow in an IP rotation. 

To minimize cable bending in an IP rotation, the rotary stage was
configured to rotate between 0 and 180$^{\circ}$. The field polarity
is reversed to cover $\varphi_{H}$ in the range between 180 and 360$^{\circ}$.
We assume that the magnetization always follows the orientation of
field ($\varphi_{H}=\varphi_{M}=\varphi$) in an IP rotation due to
the small in plane anisotropy of our samples. \ref{fig:illus_rotation}
(c) illustrates the film level measurement using a homemade multi-pin
probe. Five 45$^{\circ}$ Tungsten needles with 12$\,$$\mu$m tip
diameter were separated by 2$\,$mm. For device level measurement,
various probe types such as GSG and ground-signal (GS) RF probes with
proper pitch size can be used. \ref{fig:illus_rotation} (d) and (e)
are two examples using a GSG probe with 250 $\mu$m pitch size.

\subsection{Microwave responses}

\begin{figure*}
\centering{}\includegraphics[width=13cm]{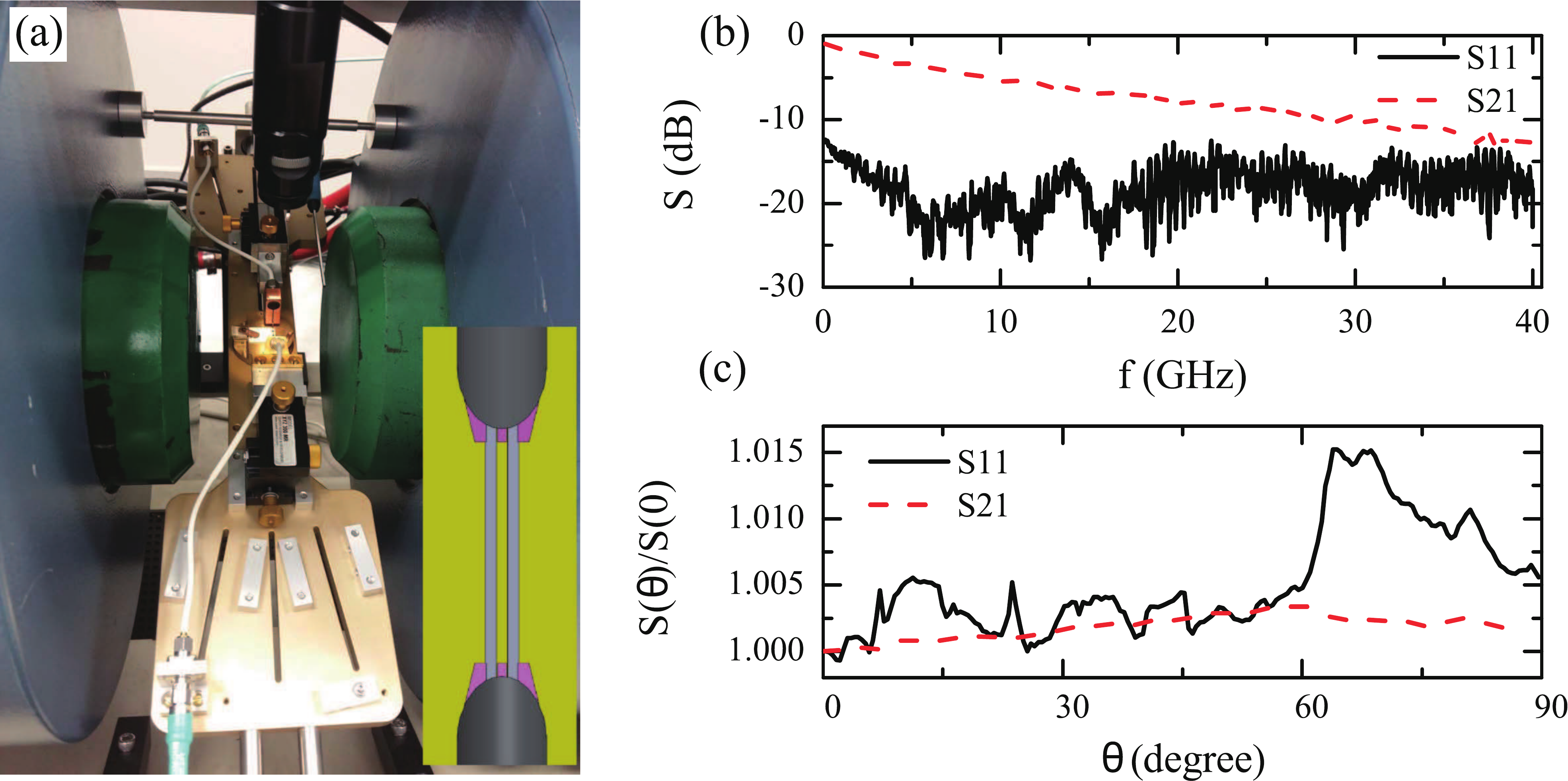}\protect\caption{Two ports scattering parameters with 0$\,$dBm input for the OP rotary
stage in an external magnetic field of 2$\,$kOe. (a) Photograph of
the measurement configuration with two GSG probes mounted. The inset
is a schematic of the probes and a calibration device. The two GSG
probes were separated by 500$\,\mu$m in a coplanar waveguide with
nominal impedance of 50 ohm. (b) Scattering parameters as a function
of frequency measured with $\theta=6\mbox{0}^{\text{\ensuremath{\circ}}}$.
(c) Normalized scattering parameters at 10$\,$GHz as a function of
$\theta$. \label{fig:circuitResponse}}
\end{figure*}

The high frequency responses of the stages including all flexible
cables were characterized by a vector network analyzer. The setup
configuration of the calibration is shown in \ref{fig:circuitResponse}
(a). Two-port scattering parameters (S11 and S21 being the reflection
coefficient and transmission coefficient, respectively) of the OP
rotary stage were measured with a 500$\,$$\mu$m long coplanar waveguide.
The microwave power is 0$\,$dBm. The magnetic field was kept at 2$\,$kOe.
Panel (b) is a plot of S11 and S21 as a function of frequency. S21
at 40 GHz is about -12.6 dB which is suitable for techniques such
as ferromagnetic resonance\citep{NTU_FMR} and pulse measurements
for most magnetic materials. The stage rotation changes the status
of cable bending, resulting in a variation in the S parameters. However,
as can be seen in \ref{fig:circuitResponse} (c), the variation in
S parameters between different orientations is normally less than
1$\%$. The frequency was fixed at 10$\,$GHz during angle sweep.
Similarly, for the IP rotary stage, S11 in ST-FMR configuration only
varies about 1$\%$ for $\varphi$ in the range between 0 and 180$^{\text{\ensuremath{\circ}}}$
(not shown). Therefore, the rotating stage method causes negligible
influence on the data analysis relevant to measurements of the angular
dependence.

\subsection{DC measurement on films and devices}

\begin{figure*}
\centering{}\includegraphics[width=13cm]{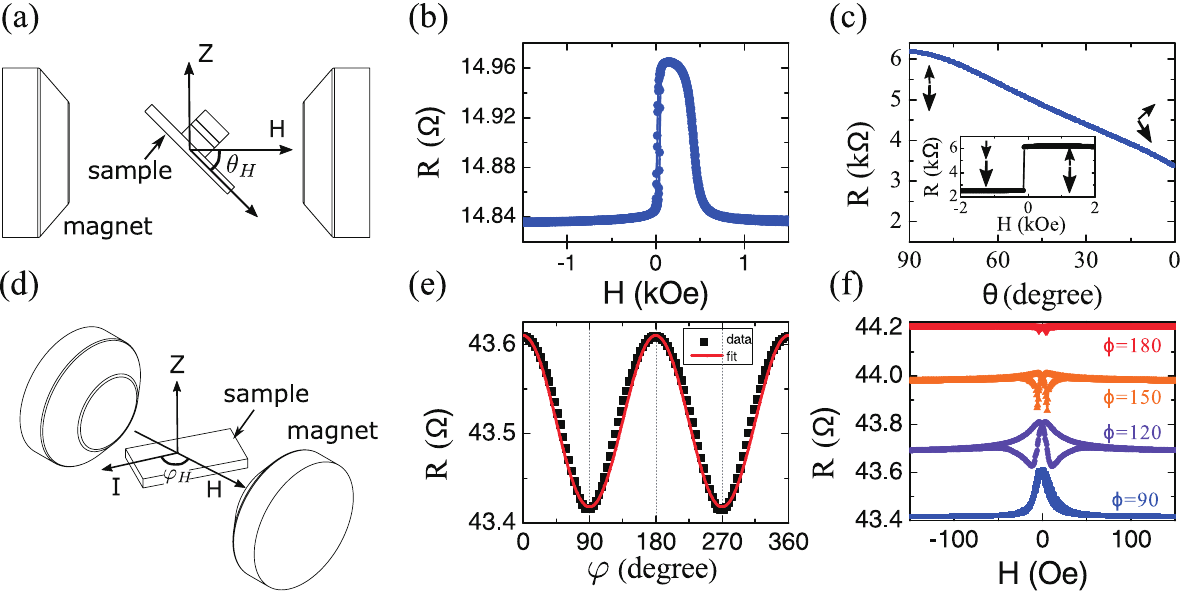}\protect\caption{(a) Side view schematic of the sample orientation using OP rotary
stage. $\theta_{\textrm{H}}$ is defined as the relative angle between
the magnetic field and the sample film plane. (b) MR loop of a Co/Cu/CoIr
based film with spin valve structure. (c) Resistance as a function
of $\theta$ for a CoFeB/MgO/CoFeB based magnetic tunneling junction
device. The field is 1.5$\,$kOe. The device is initially at high
resistance states. Inset shows the minor loop of the device. Big and
small arrows represent the magnetization orientation of the reference
and free layer, respectively. (d) Schematic view of the sample orientation
using IP rotary stage. $\varphi_{\textrm{H}}$ is defined as the relative
angle between the magnetic field and the direction of current flow.
(e) AMR of a Py/Pt stripe device measured with IP rotary stage. The
field is fixed at 1 kOe. Square and solid line are the data and the
fit to $R=R_{0}+C\cos^{2}\varphi$, respectively. (f) MR loops of
a Py/Pt device measured at selected field orientations. Curves are
shifted up by 0.2 for clarity.\label{fig:DC-results} }
\end{figure*}

For DC and low frequency measurements, a Keithley model 6221 current
source is used for current injection whereas either a Keithley model
2182A nanovoltmeter or a Stanford Research model SR830 lock in amplifier
is used to record the voltage drop across the sample. \ref{fig:DC-results}
(b) is the magnetoresistance (MR) loop taken from a film with stack
structure NiTa(150)/Pt(10)/IrMn(8)/Co (3)/Cu(3)/Co(0.5)/CoIr(3)/NiTa(5)/Pt(10)
(numbers are nominal thicknesses in nm). The setup configuration using
the OP rotary stage is illustrated in \ref{fig:DC-results} (a). The
measurement also requires four adjacent tips of the multi-pin probe
as shown in \ref{fig:illus_rotation} (c). The data was taken at $\theta$=0
with a DC current of 1$\,$mA. We observe a sudden resistance jump
close to zero field which corresponds to the free layer switch and
a broader jump around 500$\,$Oe related to the reference layer switch.\citep{fertPRL_GMR}
A similar MR minor loop on a CoFeB/MgO/CoFeB\textbf{\citep{MengHao_CEB}}
based magnetic tunnel junction (MTJ) device is shown in the inset
of \ref{fig:DC-results} (c). The loop was taken at $\theta$=90$^{\circ}$
since the device has perpendicular magnetic anisotropy.\citep{Ikeda_2010_Namat,CFB_APL_2012}
Resistance as a function of field orientation $\theta$ of the MTJ
is shown in \ref{fig:DC-results} (c). The amplitude of the field
was 1.5$\,$kOe. Note that the device was originally in the antiparallel
high resistance state at $\theta$=90$^{\circ}$. Both free layer
and reference layer were gradually tilted towards the film plane upon
changing field orientation, hence the observed decrease in the device
resistance. Tilting a free layer is necessary to optimize the performance
of a device such as a spin torque diode.\citep{STO_tiltFree2014}

Anisotropy magnetoresistance (AMR) measurement with the IP rotary
stage is also easy to perform. Shown in \ref{fig:DC-results} (e)
is an example of a 60$\,$$\mu$m$\times$30$\,$$\mu$m wide stripe
patterned on Py(6$\mbox{\,}$nm)/Pt(10$\,$nm) film. The corresponding
measurement configuration is schematically shown in \ref{fig:DC-results}
(d). The field was kept at 1$\,$kOe. Clear oscillation of the resistance
as a function of $\varphi$ was observed. The solid line is the calculated
curve using $R=43.41+0.195\cos^{2}\varphi$,\citep{IEEE_AMR_Review_1975}
where the numbers are the fitting parameters. The fit gives a AMR
ratio of $[R(0^{\circ})-R(90^{\circ})]/R(90^{\circ})$=0.45\%, smaller
than in Py films\citep{LeeAMR_Py} due to the shunting effect of Pt.
MR loops of the same device at various field orientations are shown
in \ref{fig:DC-results} (f). Sweeping the magnetic field leads to
rotation of the magnetization, which gives rise to the low-field MR
and hysteresis. The features, however, are heavily dependent on the
field direction which is indicative of magnetic anisotropy,\citep{LZU_MA_2015}
including the shape anisotropy of the rectangular stripe.

\subsection{Pulse measurement}

\begin{figure}
\centering{}\includegraphics[width=8cm]{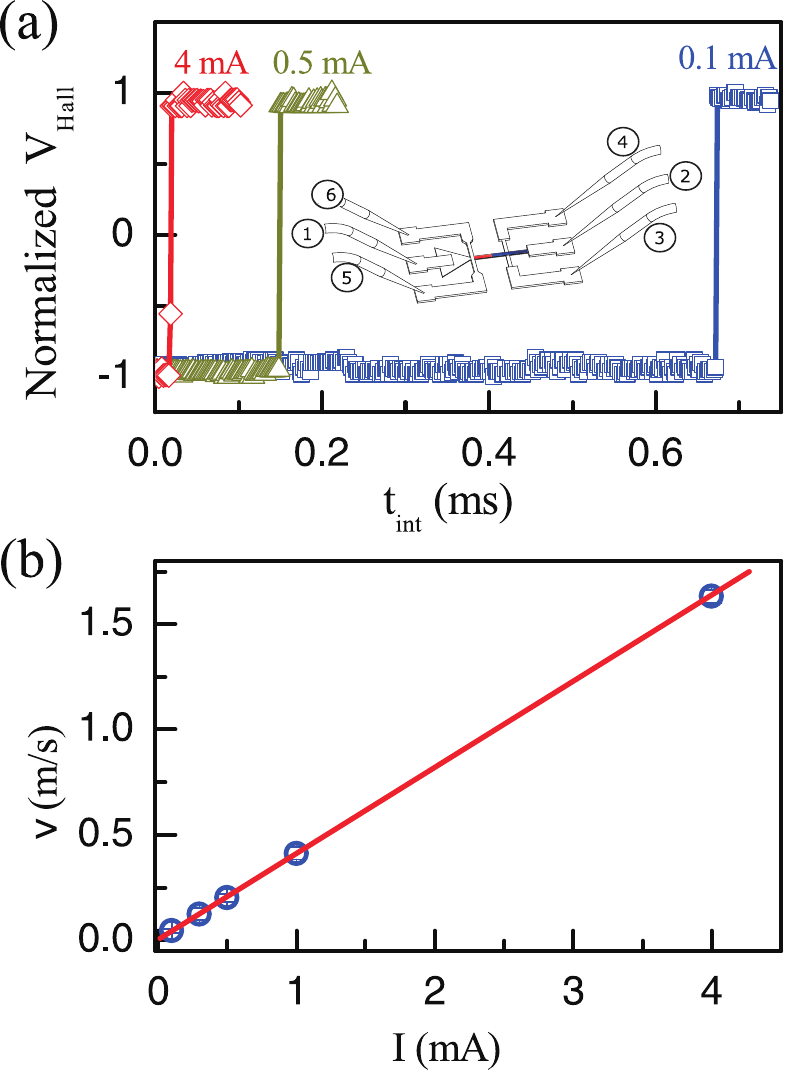}\protect\caption{Domain wall motion detection of a 300$\,$nm$\times60$$\mu$m nanowire
device patterned from a CoFe/Pd multilayer. The measurement was on
the OP rotary stage with $\theta=90^{\circ}$. (a) Hall voltage as
a function of the integrated pulse duration for selected driving current
values. The inset is a schematic of the device and multi-pin probes.
(b) The dependence of the averaged DW velocity on driving current.
Solid line is a linear fit to the data. \label{fig:DomainWall}}
\end{figure}

The bandwidth of the setup depends on the cables and probes being
used. The RF cables and probes are suitable for studying the time
dependent response of a magnetic device to pulses as short as 50$\,$ps.
On the other hand, using the homebuilt multi-pin probes, measurements
using pulses in the \textasciitilde{}ns range are also possible. A
benefit of our design is that up to 10 signal connections are available. 

The inset in \ref{fig:DomainWall} (a) shows a schematic of a domain
wall motion measurement using pulse current.\citep{Parkin_domainWall}
The stack structure of the 300$\,$nm wide nanowire device is Si/
SiO2/Ta(3)/Pd(3)/{[}CoFe(0.16)/Pd(0.22){]}8/Ta(3), where thehe number
in brackets is the thickness of the layers in nm. The nanowire was
modified with a triangular contact pad and one Hall cross for magnetotransport
measurements. In order to create a single DW in the nanowire, the
magnetization of the device was first saturated in one direction using
a large positive magnetic field (1$\,$kOe, $\theta=90^{\circ}$).
Next, a large pulse current (50$\,$mA, 50$\,$ns) was applied through
the Au electrode ( number 5 and 6) to generate a local Oersted field.
The DW was then created and pinned at the joint between the triangular
contact pad and nanowire. 

The DW motion in the nanowire was detected by means of the anomalous
Hall effect.\citep{DW_2011_NatureM} CoFe/Pd multilayers exhibit strong
perpendicular anisotropy.\citep{zhaoliang_DM} For relatively small
magnetic field values in this study, the Hall voltage is nearly proportional
to the magnetization at the Hall cross. Therefore, a sudden change
in Hall voltage is an indication of magnetization reversal due to
domain wall motion. A continuous sine-wave current was injected into
the nanowire (electrodes 1 and 2) to measure the Hall resistance through
pins 3 and 4 (inset of \ref{fig:DomainWall} (a)). Note that the amplitude
( i$_{\textrm{ac}}$=10 $\mu$A) is too small to drive the DW motion.
Two steps are repeated in the measurement. First, a 1$\,$$\mu$s
wide pulse current with desired amplitude from the pulse generator
was injected into the nanowire to drive the DW. In a second step,
the lock-in amplifier, which was locked to the frequency of i$_{\textrm{ac}}$
and with a time constant of 0.1$\,$s, was used to measure the Hall
resistance (R$_{\textrm{Hall}}$). \ref{fig:DomainWall} (a) shows
R$_{\textrm{Hall}}$ as a function of $t{}_{\textrm{int}}$ for three
different driving currents. The sudden jump in R$_{\textrm{Hall}}$
is clearly visible , indicating that the DW have passed through the
Hall cross due to the spin polarized current. The DW velocity is therefore
defined as the distance from the electrode to the Hall cross (30$\mu$m)
divided by the integrated time ($t{}_{\textrm{int}}$) of current
pulses at which the jump was observed. The extracted averaged DW velocity
($v$) as a function of driving current (I) is plotted in \ref{fig:DomainWall}
(b). Notably the DW velocity increases linearly with pulse current.

\subsection{Spin transfer torque ferromagnetic resonance}

\begin{center}
\begin{figure}
\centering{}\includegraphics[width=8cm]{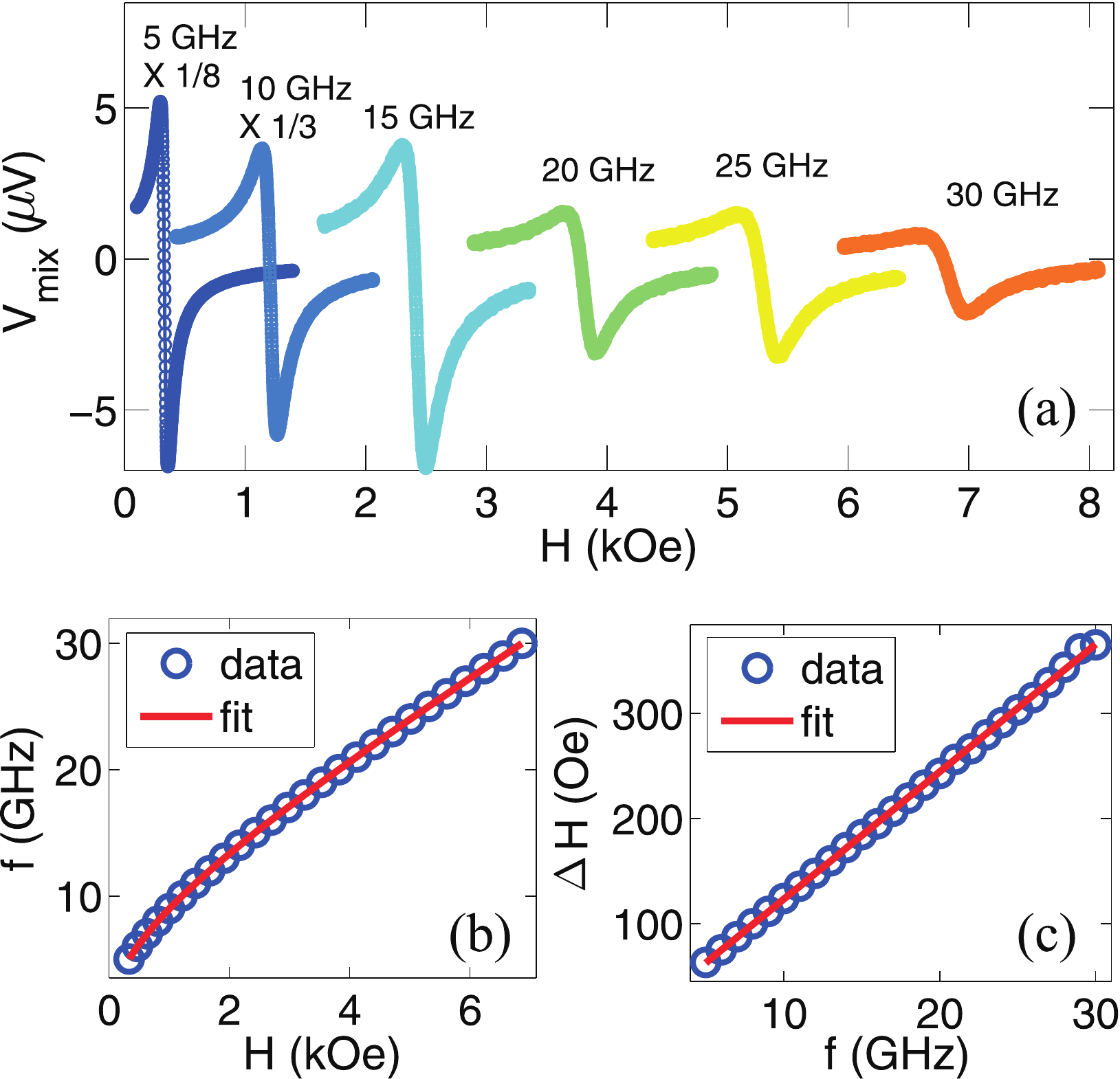}\protect\caption{ST-FMR results for a 30$\mu$m$\times$80$\mu$m Py (6$\,$nm)/Pt(10$\,$nm)
stripe taken at $\varphi=120^{\circ}$. (a) Mix voltage spectra for
selected frequencies. (b) Resonance frequency as a function of resonance
field. (c) FMR linewidth as a function of frequency. In all cases,
the circles and the solid lines correspond to the raw data and the
fit, respectively.\label{fig:ST-FMR-damping}}
\end{figure}

\par\end{center}

We have also tested our setup with spin transfer torque ferromagnetic
resonance (ST-FMR) measurement.\citep{STFMR_Liu_PRL_2011,wangyi_APL}
The device is the same Py(6$\mbox{\,}$nm)/Pt(10$\,$nm) stripe as
for \ref{fig:DC-results} (c) and (d). The device, as a load, was
integrated to a coplanar waveguide of nominal impedance 50 Ohm as
shown in \ref{fig:illus_rotation} (e). A bias Tee was used to inject
the RF current to the device and measure the DC voltage generated
in the device simultaneously.\citep{STFMR_Liu_PRL_2011} The microwave
output power was kept at 10$\,$dBm while $\varphi$=$120{}^{\circ}$.
Shown in \ref{fig:ST-FMR-damping} (a) is the mixing voltage\citep{STFMR_Liu_PRL_2011}
as a function of magnetic field for selected microwave frequencies.
Resonance peaks are clearly identified up to 30 GHz. All curves are
well fitted by the equation:\citep{STFMR_Liu_PRL_2011} 
\begin{equation}
V_{\textrm{mix}}=V_{S}\cdot F_{{\rm sym}}+V_{A}\cdot F_{{\rm asym}}\label{eq:MixVoltage_Fit}
\end{equation}
, where $F_{{\rm sym}}=(\frac{\Delta H}{2})^{2}/[(H-H_{{\rm res}})^{2}+(\frac{\Delta H}{2})^{2}]$
and $F_{{\rm asym}}=\frac{\Delta H}{2}(H-H_{{\rm res}})/[(H-H_{\textrm{res}})^{2}+(\frac{\Delta H}{2})^{2}]$
represent the symmetric and antisymmetric Lorentz component of the
resonance, respectively. $V_{S}$ and $V_{A}$ are the amplitudes
of each component. $H_{{\rm res}}$ is the resonance field and $\Delta H$
the full width at half maximum (FWHM) of a resonance spectrum. The
fitted parameters are useful for determining important material parameters
such as anisotropy and spin Hall angle.\citep{nature_TI_2014,wangyi_APL}
Shown in \ref{fig:ST-FMR-damping} (b) is the Kittel fit to the FMR
frequency as a function of resonance magnetic field. For IP magnetic
field, $f=\frac{\gamma}{2\pi}\sqrt{H_{\textrm{res}}(H_{\textrm{res}}+4\pi\textrm{\ensuremath{M_{eff}}})}$,
where $\gamma$ is the gyromagnetic ratio. The effective magnetization
4$\pi\textrm{\ensuremath{M_{eff}}}$=0.82$\pm0.01$T agrees well with
the expected value for Py.\citep{wangyi_APL} Furthermore, the linewidth
increases linearly with applied microwave frequency as shown in \ref{fig:ST-FMR-damping}
(c). The damping $\alpha=0.018$ is obtained by fitting the data to
$\Delta H=\frac{4\pi}{\gamma}\alpha f+\Delta H_{0}$. The relatively
large damping coefficient compared to that measured in Permalloy film\citep{damping_py_2007,NTU_FMR}
is attributed to the spin pumping effect due to the thick Pt layer.\citep{SpingPumping_PRL_bauer}

\begin{figure}
\centering{}\includegraphics[width=8cm]{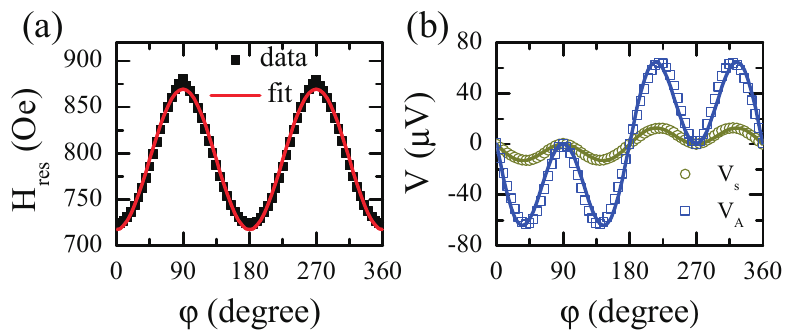}\protect\caption{Angular dependence of the ST-FMR for a Py (6$\,$nm)/Pt(10$\,$nm)
stripe measured in the IP rotary stage. (a) The resonance field as
a function of $\varphi$ at $f=8\mbox{\,}$GHZ. The solid line is
the calculated curve using \ref{eq:IPR_kittel} with $4\pi M_{{\rm eff}}$=0.84$\pm0.01$$\,$T
and $H_{2\parallel}=79\pm2$$\,$Oe. (b) Symmetric and antisymmetric
component of ST-FMR spectra as a function of $\varphi$. Solid lines
are the fit to a $\textrm{cos}{}^{2}\varphi\textrm{sin}\varphi$ function.\label{fig:IPR_STFMR}}
\end{figure}

Compared to the measurement at a fixed angle, we are more interested
in the angular dependence of the ST-FMR spectrum. \ref{fig:IPR_STFMR}
shows the IP angular dependence of the resonance field and resonance
amplitudes measured at 8$\,$GHz. The data points are extracted by
fitting 72 spectra collected at a list of field orientations. For
this particular sample, besides the perpendicular uniaxial anisotropy
which is included in $4\pi M_{{\rm eff}}$, an additional in-plane
uniaxial anisotropy (effective field $H_{2\parallel}$) with easy
axis along the long-side is assumed. Using the coordinates defined
in \ref{fig:illus_rotation}, the resonance condition for any given
IP field orientation can be derived:\citep{1973_FMR_angular,Liu_angular_GaAs} 

\begin{widetext}

\begin{equation}
2\pi f=\gamma\sqrt{(H_{{\rm res}}+4\pi M_{{\rm eff}}+H_{2\parallel}\cos^{2}\varphi_{M})(H_{{\rm res}}+H_{2\parallel}\cos2\varphi_{M})}\label{eq:IPR_kittel}
\end{equation}

\end{widetext}

Thus, $4\pi M_{{\rm eff}}$=0.84$\pm0.01$T and $H_{2\parallel}$=62$\pm$2
Oe . Py/Pt is a well investigated bilayer for the spin Hall effect
(SHE).\citep{RMP_2015_SHE} The antisymmetric and symmetric voltage
components are attributed to field-like torque and spin torque generated
by the SHE, respectively. The output DC voltages arise from the combined
effect of AMR and magnetization precession driven by these torques.\citep{mixVoltage_formula_Azevedo}
A $\textrm{cos}{}^{2}\varphi\textrm{sin}\varphi$ relation is expected
if no other effect is involved.\citep{STFMR_IPR_harder2011,IPR_symmetry_APL2014,PhysRevB_IPR_CMHu}
As can be seen in \ref{fig:IPR_STFMR} (b), both components can be
fitted well by this relation, adding credence to the suitability of
the IP rotary stage for angular dependent studies. Note that the measurement
only takes 4$\,$h with a single program sequence. In contrast, using
probe stations equipped with rotary sample stages, each data point
shown in \ref{fig:IPR_STFMR} requires a separate probe landing and
stage rotating procedures. It can take several days to obtain the
same curve. Therefore, the method of rotating sample and probes simultaneously
is much more efficient.

\section{conclusion}

We have developed a probe station to study magnetic films and devices
with controllable field orientation. The sample holder, probe assembly
and microwave cables are rotated simultaneously with two separate
motorized stages. The sample rotation is equivalent to a field rotation
which covers $\theta$ from 0 to $90^{\circ}$ and $\varphi$ from
0 to $360^{\circ}$. We have also shown novel designs of a vision
system assembly, sample holder with spring-loaded clip, micro positioner
mounting and multi-pin probe. Furthermore, a range of DC and RF measurement
techniques incorporated to this probe station were employed to demonstrate
the versatility of this probe station. The angular-resolved RF probe
station is useful for determining magnetic anisotropy, evaluating
device performance and addressing the physics of spin torque.
\begin{acknowledgments}
The authors are grateful to Yi Wang, Sze Ter Lim, Cheow Hin Sim and
Ruisheng Liu for fruitful discussions. We thank HongJing Chung for
preparing films. We acknowledge Singapore Ministry of Education (MOE),
Academic Research Fund Tier 2 (Reference No: MOE2014-T2-1-050) and
National Research Foundation (NRF) of Singapore, NRF-Investigatorship
(Reference No: NRF-NRFI2015-04) for the funding of this research.
\end{acknowledgments}

\bibliographystyle{apsrev4-1}
\bibliography{RSI-DSI-probestation}

\end{document}